\documentclass[aps,prl,longbibliography,twocolumn,superscriptaddress,showkeys,floatfix]{revtex4-2}

\usepackage{graphicx}
\usepackage{dcolumn}
\usepackage{bm}
\usepackage{xcolor}
\usepackage{float}  
\usepackage{hyperref}
\usepackage{amsfonts} 
\usepackage{amsmath} 
\hypersetup{  
	colorlinks,
	linkcolor={red!50!black},
	citecolor={blue!50!black},
	urlcolor={blue!80!black}
}

\begin{document}

\title{Flexibility in noisy cell-to-cell information dynamics}

\author{Ismail Qunbar}
\thanks{These authors contributed equally to this work.}
\affiliation{%
 Racah Institute of Physics,
 The Hebrew University of Jerusalem,
 Jerusalem 9190401, Israel
}%
\author{Michael Vennettilli}
\thanks{These authors contributed equally to this work.}
\affiliation{%
 AMOLF, 
 Science Park 104, 1098 XG Amsterdam, 
 The Netherlands
}%

\author{Amir Erez}
\email{amir.erez1@mail.huji.ac.il}
\affiliation{%
 Racah Institute of Physics,
 The Hebrew University of Jerusalem,
 Jerusalem 9190401, Israel
}%

\date{\today}

\begin{abstract}
Exchange of molecules allows cells to exchange information. How robust is the information to changes in cell parameters? We use a mapping between the stochastic dynamics of two cells sharing a stimulatory molecule, and parameters akin to an extension of Landau's equilibrium phase transition theory. We show that different single-cell dynamics lead to the same dynamical response---a flexibility that cells can use. The companion equilibrium Landau model behaves similarly, thereby describing the dynamics of information in a broad class of models with coupled order parameters.
\end{abstract}

\maketitle

In the choreography of life, cells respond to environmental stimuli by detecting chemical signals and secreting molecules. This process, fundamental to cellular communication, has attracted much attention, and recent advances in our understanding of information transfer in biological systems include: quantifying the energy cost of information transmission in coupled receptors \cite{ngampruetikorn2020};  interpreting positional information and spatial coupling \cite{sokolowski2015, tkacik2021}; establishing bounds on the energy cost of transmitting information in various settings \cite{bryant2023}; outlining essential trade-offs between cost and predictive power \cite{tjalma2023}. These studies underscore the complexity and efficiency of cellular communication using molecular exchange and highlight the importance of understanding how cells manage and optimize information exchange. However, despite the dynamic nature of cellular environments, a scaling theory for cell-to-cell information dynamics under changing conditions had yet to be developed.

Our previous work described a minimal model of feedback in cellular sense-and-secrete dynamics \cite{erez2019} and investigated how pairs of such cells share information \cite{erez2020}. For example, upon antigen stimulation, T cells exhibit a bimodal distribution of doubly phosphorylated ERK (ppERK), a critical protein that initiates cell proliferation and determines the immune response \cite{vogel2016,erez2018, altan-bonnet2005}. We were able to show that a broad class of models, including Sch\"ogl's second model, can be used to explain and extract key features from single cell measurements of ppERK \cite{erez2019,byrd2019}. Here, we extend this line of inquiry to explore how time-varying individual cellular properties impact overall information exchange. How does information in a minimal sense-and-secrete two-cell system respond to gradually changing cellular parameters?
We demonstrate that inherent flexibility allows one cell to match the dynamics of the other, in a choreographed dance, so that in a broad class of models, cell-to-cell information is maintained and key observables scale universally.

Our minimal model of cellular sense-and-secrete dynamics maps a class of well-mixed stochastic biochemical feedback models, in steady state, to parameters analogous to those in Landau’s thermal equilibrium phase transition theory. Though applicable to other stochastic models, we focused specifically on mapping the dynamics of Schl\"ogl's second model \cite{schlogl1972,grassberger1982} as if it were a Landau theory. Instead of stochastic reaction rates, one may describe the dynamics with an effective reduced temperature, $\theta$, magnetic field, $h$, and a magnetization-like order parameter, $m$ \cite{erez2019, erez2020}. These parameters can be extracted from biological data without fitting or knowledge of the underlying molecular details \cite{erez2019}. The stochastic dynamics are never in thermal equilibrium, and the noise is demographic in nature. For this reason we will refer to the Sch\"ogl model at steady state and the Landau model at thermal equilibrium as `companion' systems. The critical transition from a finite to a zero `magnetization' in the Landau theory is equivalent to the bifurcation point of the stochastic dynamics \cite{bose2019}---between a bimodal state, having both high and low molecule counts as stable points the dynamics fluctuate about---to a unimodal state with an intermediate molecule count (\emph{cf}. SI Appendix Sec.~A). Near the transition point, the correlation time of the system diverges, exhibiting critical slowing down as expected from the Landau theory \cite{byrd2019}. 

A diverging timescale and critical slowing down do not matter much when considering systems at equilibrium or steady state. In contrast, in a dynamically changing setting, collective properties are expected to be influenced by the slowing down of the collective dynamics. The two-cell system (cells: $X$ and $Y$) has four main parameters: two `fields' ($h_X, h_Y$) that control overall molecule count bias in each cell and two `reduced temperatures' ($\theta_X, \theta_Y$) that act as a bifurcation parameter. The steady state can be described by two collective coordinates: $H,T$. These collective coordinates dictate both the steady state mutual information between the cells and the autocorrelation time of the molecule count, which are closely related \cite{erez2020,vennettilli2020}. In this two-cell collective state, each individual cell's reaction rates can be set away from its critical point, while the two-cell collective system remains at criticality. Because the collective state is critical, both the mutual information and the correlation time are maximized; to gain information at steady state one must `pay' with an increased correlation time \cite{erez2020}. Importantly, the collective coordinates $H,T$ are steady-state, linearized solutions that need not apply to the system's dynamics. Here, we consider the univesal properties of the two-cell response to time-varying conditions.

Can the system's dynamical changes be described using the collective coordinates, disregarding each cell's specific dynamics? In this manuscript, we demonstrate that they can be. Therefore, diverse single-cell parameter trajectories can lead to identical systemic responses, allowing substantial flexibility in single-cell properties without impacting the two-cell information. This flexibility in single-cell behavior enhances robustness and evolutionary adaptability, facilitating the exploration of advantageous strategies. Furthermore, while our focus is on cell-to-cell communication, manifested through the stochastic non-equilibrium dynamics of the Sch\"ogl model, our study of the companion Landau model reveals a similar degeneracy in the analogous equilibrium system. This extension of equilibrium Landau theory has been used in the context of phase transitions in minerals and surface and hydration forces \cite{salje1992,kornyshev1992,kornyshev1995}. Therefore, this study describes a general class of models with coupled order parameters, both equilibrium and non-equilibrium, that bifurcate according to the Ising mean-field universality class. 

The rest of this manuscript is organized as follows: after a brief recap of the model, we compare the steady state of the stochastic dynamics of the extended Schl\"ogl  model to its equilibrium Landau companion. We then dynamically ramp both cells' parameters across the bifurcation or critical point, showing the lag in the dynamical response of the ramped system and its scaling. Finally, we show how different single-cell trajectories, sharing the same collective coordinates, lead to identical responses in the information exchange between the two cells.

\begin{figure}[ht]
    \centering
    \includegraphics[width=0.99\columnwidth]{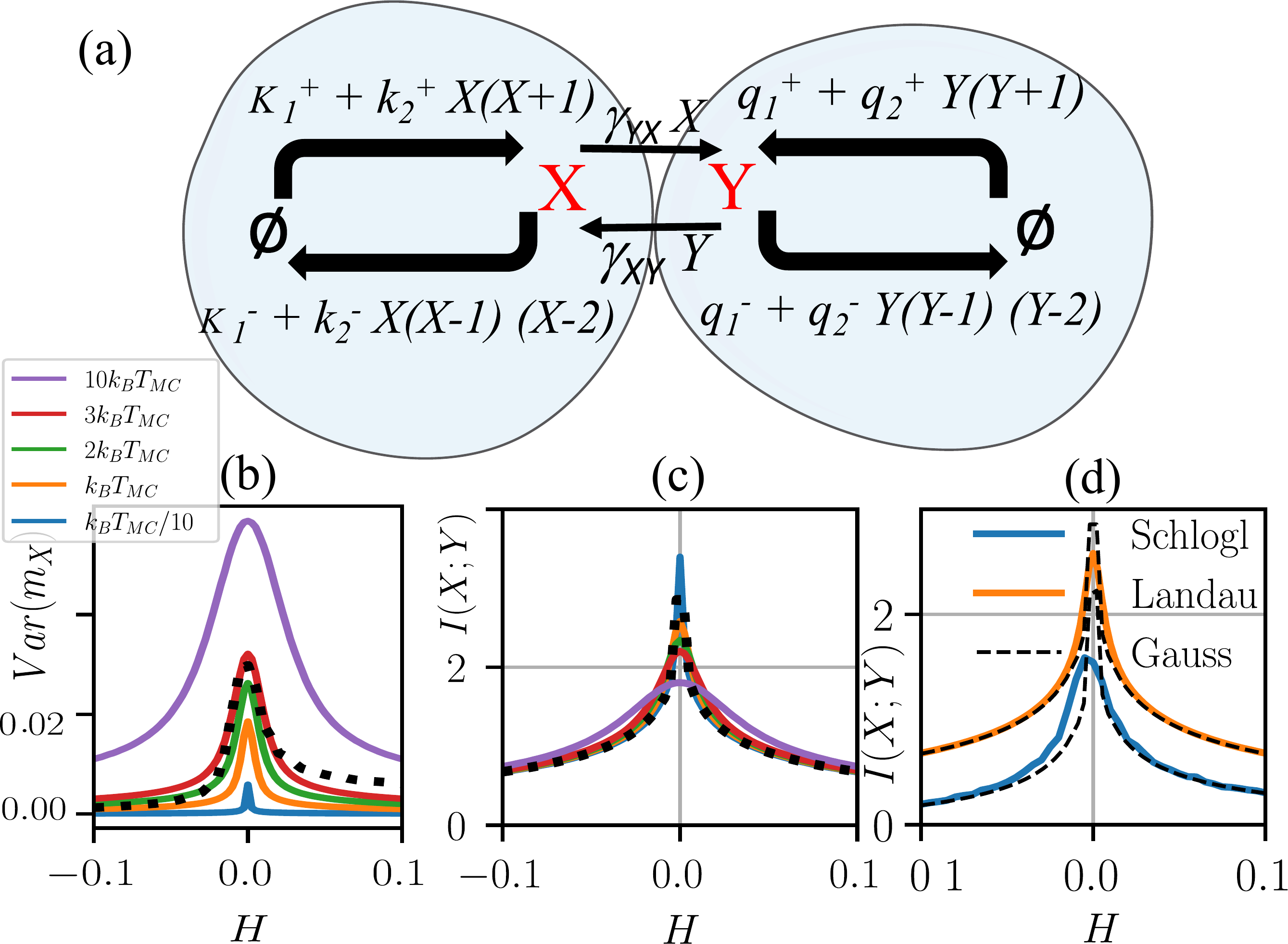} \\
     \caption{Schematic of the two-cell model and its steady state. (a) Representation of the two-cell version of the stochastic dynamics, which extend Schl\"ogl's second model \cite{schlogl1972}, similarly to our previous manuscript \cite{erez2020}. (b) Varying the Monte-Carlo temperature $k_B T_{MC}$ in the Landau model affects some observables, e.g., the variance of $m_X$. The Schl\"ogl model (black dots) exhibits a dependence on $H$ which cannot be captured by tuning the Landau model's $k_BT_{MC}$, reflecting its non-equilibrium nature. (c) The mutual information in the Landau model depends only weakly on $k_BT_{MC}$, and closely matches the Gaussian analytic result (black dots), though at $T=H=0$ the mutual information in the Gaussian case diverges. (d) Comparison of the mutual information of the two-cell system as a function of the collective field, $H$. Blue---the non-equilibrium Sch\"ogl steady state; Orange---the extended Landau model at thermodynamic equilibrium. Dashed black---the analytic results from the Gaussian approximation. Here, both cells are identical: $h_x=h_y=h$ and $\theta_x=\theta_y=0$, with $n_c=1000$.} 
    \label{fig:intro}
\end{figure}

\emph{Model and mapping to Landau's equilibrium phase transition theory}---Within each cell, biochemical reactions in a complex signaling cascade result in the net production and degradation of a molecular species of interest. As illustrated in Fig.~\ref{fig:intro}a, in the first (second) cell, species $X$ ($Y$) can be produced spontaneously from bath species at rate $k^+_1$ ($q^+_1$), and can be produced nonlinearly at rate $k^+_2$ ($q^+_2$) via a trimolecular reaction involving two existing $X$ ($Y$) species and a bath species. Species $X$ ($Y$) can be degraded linearly with molecule number at a rate $k^-_1$ ($q^-_1$), or in a reaction involving three existing $X$ ($Y$) molecules at rate $k^-_2$ ($q^-_2$). In addition to the internal reactions, $X$ ($Y$) can be exchanged from the neighboring cell at rate $\gamma_{xy}$ ($\gamma_{yx}$). Mechanistically, this can be through a gap junction or through diffusion \cite{erez2020}. 

Reiterating our previous work \cite{erez2019, byrd2019}, we use a mapping from Schl\"ogl to Landau-like parameters. Without exchange, ($\gamma=0$), the deterministic dynamics corresponding to the reactions in the left cell in Fig.~\ref{fig:intro}a are $dx/dt = k_1^+ - k_1^-x + k_2^+x^2 - k_2^-x^3$, where we have neglected the small shifts of $-1$ and $-2$ for large $x$. Defining the order parameter $m = (x-n_c)/n_c$, we choose $n_c$ to eliminate the term quadratic in $m$, leading to the Landau form \cite{erez2019},
\begin{equation}
	\label{eq:landau}
	\frac{dm}{d\tau}=h-\theta m -\frac{m^3}{3},
\end{equation}
where we have defined $n_c = k_2^+/3k_2^-$, $\tau = (k_2^+)^2t/3k_2^-$, $\theta = 3k_1^-k_2^-/(k_2^+)^2 - 1$, and $h = 9k_1^+(k_2^-)^2/(k_2^+)^3 - 3k_1^-k_2^-/(k_2^+)^2 + 2/3$. The number of molecules in the system is controlled by $n_c$, which controls all scaling properties of the single-cell system, acting as a \emph{finite system size} of the equivalent critical Ising system \cite{erez2019,byrd2019}. 

In steady state, $dm/d \tau=0$. We can thus interpret $m$ as an order parameter for the single-cell system, $\theta \equiv (T-T_c)/T_c$ as a `reduced temperature', and $h$ as a dimensionless field. Analogous to the Ising model, when $h=0$ in the single-cell system, $\theta>0$ corresponds to a unimodal steady-state distribution, and $\theta<0$ to a bimodal distribution. Similarly, tuning $h$ biases the distribution to high or low molecule count. Applying the same mapping to two coupled cells (with $k\to q$ for $Y$) results in the Landau form,
\begin{eqnarray}
		\label{eq:LandauForm}
		\frac{dm_X}{d\tau}&=&h_X-\theta_X m_X - \frac{m_X^3}{3}+g_{XY}m_Y- g_{YX}m_X\,, \nonumber \\
		\frac{dm_Y}{d\tau}&=&h_Y-\theta_Y m_Y - \frac{m_Y^3}{3}+g_{YX}m_X-g_{XY}m_Y\,.
\end{eqnarray}
In this context, $g_{XY}=3\gamma_{xy}k_2^-/(k_2^+)^2$ and $g_{YX}=3\gamma_{yx}q_2^-/(q_2^+)^2$ represent intercellular exchange terms, set to unity throughout this manuscript. 

Although Eq.~\ref{eq:LandauForm} is a reparameterization of the nonequilibrium stochastic dynamics of the extended Sch\"ogl model (Fig.~\ref{fig:intro}a), one may separately consider it as an extension of the well-known Landau model (Eq.~\ref{eq:landau}), computed at thermal equilibrium. This extension of equilibrium Landau theory to two bilinearly-coupled order parameters has been considered before: in the context of phase transitions in minerals \cite{salje1992}; surface and hydration forces \cite{kornyshev1992}; and more generally \cite{kornyshev1995}. Here we focus on the  biological system, with its non-equilibrium stochastic dynamics, while in parallel demonstrating similar dynamical scaling results for the companion Landau theory computed at thermal equilibrium.

Linearizing the deterministic steady state ($\frac{d}{d\tau}=0$) of the Landau form (Eq.~\ref{eq:LandauForm}) gives the collective coordinates,
\begin{eqnarray}
	\label{eq:HT}
	T &=& \theta_X\theta_Y+ g(\theta_X+\theta_Y)\,, \nonumber\\
	H &=& g(h_X+h_Y) + \left(  h_X \theta_Y +h_Y \theta_X\right)/2\,.
\end{eqnarray} 
At the critical point, $T=H=0$, the two cells exhibit critical slowing down and maximal mutual information between $X$ and $Y$, regardless of each cell's individual $h,\theta$ values \cite{erez2020}. 
	
\emph{Results}---We simulated the Sch\"ogl system's stochastic dynamics using the Gillespie algorithm \cite{gillespie1976,gillespie2000}, and the equilibrium fluctuations of the companion Landau model with the Metropolis–Hastings Monte Carlo algorithm. The dynamics in the Landau case are taken as `model A' in the Halperin-Hohenberg classification \cite{hohenberg1977, goldenfeld1992},
\begin{equation}
    \frac{dm_X}{d\tau} = -\Gamma \frac{\delta L}{\delta m_X} + \zeta(\tau)\,,
\end{equation}
and similarly for $m_Y$, with $\Gamma$ setting the relaxation timescale of the system and $L = -h_X m_X + \frac{1}{2} \theta_X m_X^2 +\frac{1} {12}m_X^4 -h_Y m_Y + \frac{1}{2} \theta_Y m_Y^2 +\frac{1} {12}m_Y^4 +\frac{1}{2}\,g\,(m_X-m_Y)^2$. The noise, $\zeta(t)$, is Gaussian white noise obeying $\langle \zeta \rangle = 0$ and $\langle \zeta(\tau) \zeta(\tau') \rangle = \delta(\tau-\tau') D$. In the equilibrium statistical mechanics sense, to calculate the partition function, or alternatively to compute expectation values of observables using the Monte Carlo simulation accept/reject, we require a simulation temperature, which we refer to as $k_BT_{MC}$. The long-time limit of the dynamics as a Fokker-Planck equation approaches the equilibrium solution with $P(m_X,m_Y)\propto \exp\left(-\frac{2\Gamma L}{D}\right)$ and accordingly a fluctuation-dissipation relation, $D=2\Gamma k_BT_{MC}$ \cite{goldenfeld1992}. To set the equilibrium temperature $k_BT_{MC}$ we resort to a heuristic argument: near the bifurcation point of the dynamics, ($\theta=0, h=0$), we have $\mbox{mean}[X]\sim\mbox{Var}[X]\sim n_c$ and therefore $\mbox{Var}[m_X]=\mbox{Var}[X]/n_c^2\sim 1/n_c$. Away from $H=0$ but at $\vert H\vert \ll g$, the Gaussian approximation gives $\mbox{Var}[m_X]\sim k_BT_{MC}$ (\emph{cf.} SI Appendix Eq.~\ref{eq:varM}). Therefore, $k_B T_c \sim 1/n_c$. However, tuning the precise value of $k_BT_{MC}$ in comparison to the stochastic simulations of the Sch\"ogl model is pointless since the steady state of the Sch\"ogl model does not comply with the equilibrium form, $\exp\left(-\frac{2\Gamma L}{D}\right)$. Therefore, the probability distribution of $m_X$, and an observable such as the variance, $\mbox{Var}[m_X]$, depend on $k_BT_{MC}$ in the thermalized Landau model (Fig.~\ref{fig:intro}b, colored curves). No value of $k_BT_{MC}$ in the Landau simulation will truly capture the nonequilibrium Sch\"ogl steady state (Fig.~\ref{fig:intro}b, dashed black). In this manuscript, setting the precise value of $k_B T_{MC}$ is not necessary since we focus on the mutual information between the two cells, a quantity that changes only very weakly with $k_BT_{MC}$ (Fig.~\ref{fig:intro}c, colored curves). 

The mutual information shown in Fig.~\ref{fig:intro}c is very well approximated by a Gaussian analytical approximation (dashed black). The approximation linearizes the dynamics around the deterministic fixed point: whether of the Landau model or the chemical Langevin description of the Schl\"ogl model. This yields a multi-dimensional Ornstein-Uhlenbeck process \cite{klebaner2012}: $
    d\vec{X}_t = \mathcal{J}\left(\vec{X}_t - \vec{\mu} \right)dt + \sigma d\vec{W}_t\,,$
where $\vec{X}$ is the column vector of variables of interest, magnetizations or molecule numbers, $\mathcal{J}$ is a negative-definite matrix, $\vec{W}$ is a vector of independent Brownian motions, and $\sigma$ is a matrix. The solution is a Gaussian process, where the mutual information is completely determined by the covariance matrix. By computing the pathwise solution (\emph{cf.} SI Appendix Sec.~B and \cite{klebaner2012}), one finds that the steady state covariance matrix $\mathcal{C}$ satisfies the Lyapunov equation
$
    \mathcal{J}\mathcal{C} + \mathcal{C}\mathcal{J}^T = -\sigma\sigma^T$.
This is a linear system for the coefficients of $\mathcal{C}$, so all that is left is to determine the matrices $\mathcal{J}$ and $\sigma$ for the two models. For the Landau model with $0< |H| \ll g$, we find
    $I_{\text{Landau}} = -\frac{1}{2} \log\left(1 - \rho^2\right)$, with $\rho^2 = \frac{g^2}{(g+\overline{m}^2)^2}$
and $\overline{m} =\left(3H/2\right)^{1/3}$. Under the Gaussian approximation, the covariance matrix is proportional to $k_B T_{MC}$, so the temperature cancels out when computing the correlation coefficient. Therefore, to Gaussian order, the mutual information is insensitive to the choice of $k_B T_{MC}$. Since $I_{\text{Landau}}$ is an even function of $\overline{m}$, the information is symmetric in $H$. Repeating the process for the Schl\"ogl model, restricting to $h_X = h_Y$, we find $I_{\text{Schl\"ogl}} = -\frac{1}{2} \log\left(1-\rho^2 \right)$, with $\rho^2=\frac{4 g^2 \left(-\overline{m}^2+\overline{m}+2\right)^2}{\left(2 g \left(2
   \overline{m}^2+\overline{m}+2\right)+\left(\overline{m} \left(\overline{m}+2\right)+4\right)
   \overline{m}^2\right)^2}$.
This Gaussian approximation for the Schl\"ogl model is not even in $\overline{m}$, and therefore asymmetric in $H$. As $H\rightarrow 0$, the Gaussian information diverges, indicating that the cubic terms are needed for stabilization and suggesting that the mutual information attains a maximum. For the mathematical details, \emph{cf.}  SI Appendix Sec.~B and \cite{klebaner2012}.

Setting $\theta_x=\theta_y=T=0$ and $h_x=h_y$, and therefore $H=h_x+h_y$, we calculated numerically the mutual information between $X$ and $Y$ at steady state for the Sch\"ogl model and at equilibrium for the Landau case. As expected, the mutual information is maximized at the critical point, $H=0$ (Fig.~\ref{fig:intro}d). The Gaussian approximation captures well the numerical simulations away from the critical point, including the assymetry in the Schl\"ogl case, reflecting the slightly higher information at positive $H$ due to a higher molecule count.  Interestingly, despite the typical number of molecules, $n_c$, being the same in both systems, the mutual information for the equilibrium Landau model is higher than the companion Schl\"ogl model. 

When a system is gradually driven through a critical point, critical slowing down causes a lagged response to the driving. This phenomenon, known as the Kibble-Zurek (KZ) effect in the statistical physics literature, results in a lag that follows scaling rules governed by the critical exponents of the transition point \cite{kibble1976,zurek1985,biroli2010,chandran2012, deffner2017, fujita2020, proverbio2022, tarantelli2022, berger2023}. Does our two-cell system exhibit KZ scaling when the collective coordinate $H$ is driven across the transition point? While various driving protocols are possible in a biological context, when crossing the critical point, terms beyond the leading-order linear term do not asymptotically alter the critical scaling \cite{chandran2012}. This theoretical advantage allows us to focus on a simple protocol of $H(t)$ changing linearly with time without losing biological realism---any reasonable trajectory with $H(t)$ crossing the transition point at a constant rate should scale the same.

\begin{figure}[ht]
	\centering
	\vspace{0.01\textheight}
	\includegraphics[width=0.99\columnwidth]{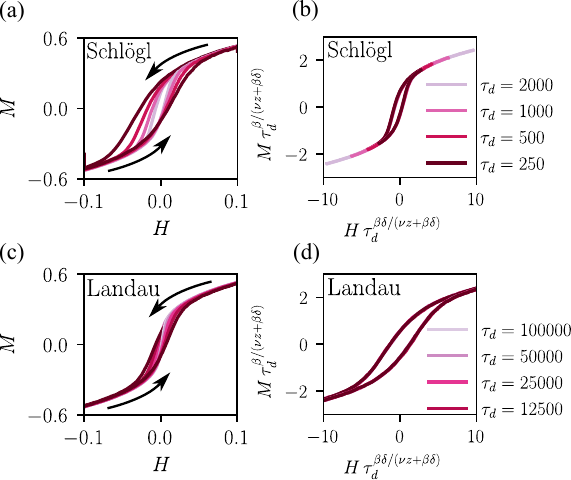}
	\caption{Lagged response of the collective magnetization of the two-cell system, $M=(m_X+m_Y)/2$, to time-varying $H(t)$ as it is driven across the transition point at $H=0$. The collective field, $H(t)$ is ramped according to $H(t)=H_i + (H_f-H_i)/\tau_d$. The arrows indicate the direction of the ramp, starting at either $H_i=0.1$ or $-0.1$. (a) Schl\"ogl dynamics for $\tau_d=2000, 1000, 500, 250$. To control for finite molecule-number effects, we scale $n_c = 4\, \tau_d^{4/5}$ \cite{chandran2012,byrd2019}. (b) The Schl\"ogl results scaled according to the Kibble-Zurek scaling prediction. (c) The extended Landau system with $\tau_d=100000, 50000, 25000, 12500$ and $n_c =\frac{1}{5}\, \tau_d^{4/5}$. (d) Kibble-Zurek collapse of the Landau system response. In all figures, as in the Ising mean-field universality class, $\beta=\frac{1}{2}, \nu z = 1, \delta=3$.}
  \label{fig:KZ_samepath}
\end{figure}

\emph{Hysteresis in the magnetization}---Maintaining $T=0$, we first let the system relax at $H=H_i=0.1$, then varied $H(t)$ from $H_i$ to $H_f=-0.1$ at a rate determined by $\tau_d$, i.e., $H(t)=H_i + (H_f-H_i)/\tau_d$. Similarly, we initiated at $H_i=-0.1$ and reversed the trajectory. Importantly, to control for the damping effect of finite-molecule number on the divergence of the correlations, the system size, $n_c$, must be scaled according to $n_c\sim \tau_d^{4/5}$ \cite{chandran2012,byrd2019}. Consistent with KZ theory, driving across the transition induced hysteresis loops in both the Schlögl and Landau cases, with loop characteristics dependent on the ramp time, $\tau_d$, and the critical exponents of the transition point. In both cases when scaled as per KZ predictions, $M\tau_d^{\beta/(\nu z+\beta \delta)}\sim H\tau_d^{\beta\delta/(\nu z+\beta \delta)}$, the Ising mean-field values for the exponents led to the collapse of the disparate hysteresis loops to a single curve  (Fig.~\ref{fig:KZ_samepath}).   

\emph{Flexibility in dynamics}---There are many different trajectories in $\{h_X,h_Y,\theta_X,\theta_Y\}$ that have the same values for collective $\{H,T\}$. We wanted to discern whether the dynamical response of the system is sensitive only to $\{H,T\}$, even when their derivation assumed steady state. To compare the dynamical response of the system to different trajectories, we considered Eq.~\ref{eq:HT} with $H(t) = H_i + (H_f-H_i)\,t/\tau_d$ and $T=0$, setting $H_i=0.1$ and $H_f=-0.1$, thereby crossing the two-cell transition point $H=T=0$. To maintain $T=0$, we set  $\theta_Y=-\frac{\theta_X}{\theta_X+1}$. Therefore, in terms of $H(t)$, the $\{h_X(t),h_Y(t)\}$ trajectories must obey $h_Y(t)=\frac{1}{1-\theta_Y/2}\left[H(t) - h_X(t) \left(1-\frac{\theta_X/2}{\theta_X+1}\right)\right]$.
For the special case of $\theta_X=\theta_Y=0$ we retrieve $h_Y(t) = H(t) - h_X(t)$, where we considered two situations: $h_Y(t)=h_X(t)$ as in Fig.~\ref{fig:KZ_samepath}, and $h_Y(t)=h_X(t)-0.1$. Furthermore, we considered a third trajectory where $\theta_x(t)=0.04 - 0.08 (t/\tau_d)$ and $h_x(t) = 0.1-0.2 (t/\tau)^2$. We chose the quadratic dependence on time to make explicit the freedom in choosing the trajectories of two of the four control parameters while maintaining the same collective dynamics. Though many other trajectories are possible, we trust that demonstrating that these three examples give identical information is enough to establish the generic flexibility in the cell-to-cell dynamics. 

\begin{figure}[ht]
    \centering
    \includegraphics[width=0.99\columnwidth]{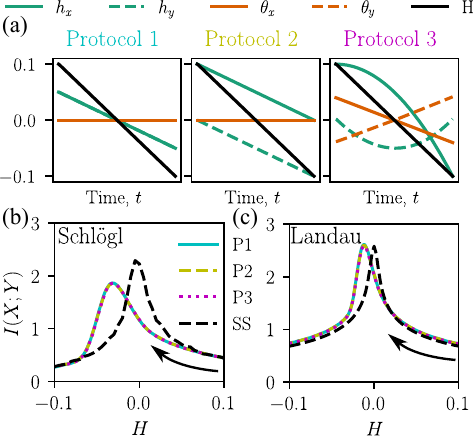}
    \caption{Mutual information for the two-cell system with dynamically changing parameters. (a) Three different driving protocols are shown, with the same collective dynamics, $T=0$ and $H(t)=H_i+(H_f-H_i)/\tau_d$, but different realizations of $\{h_x(t), h_y(t), \theta_x(t), \theta_y(t)\}$. Protocol 1: $h_x(t)=h_y(t)$ and $\theta_x=\theta_y=0$; Protocol 2: $h_x(t) = 0.1+h_y(t)$ and $\theta_x=\theta_y=0$; Protocol 3: $\theta_x(t)=0.04 - 0.08 (t/\tau_d)$ and $h_x(t) = 0.1-0.2 (t/\tau)^2$ (see text). (b) Mutual information calculated from the stochastic dynamics of the Schl\"ogl model, demonstrating the same behavior of all three protocols (P1, P2, P3), with $\tau_d=250$, showing lag and hysteresis when the transition point is crossed. Dashes: the steady-state mutual information from Fig.~\ref{fig:intro}d. The black arrow indicates the direction of the parameter ramp. (c) Mutual information of the companion Landau model simulated using Metropolis Monte Carlo, with $\tau_d=6400$, showing convergence of the three protocols. Dashes: the equilibrium mutual information. In all simulations, $n_c=1000$.}
    \label{fig:KZ_twocell}
\end{figure}

Contrary to the steady-state calculation (Fig.~\ref{fig:intro}) where one tracks the system over a sufficiently long time---here, to compute the information at each time-step we averaged over an ensemble of independent stochastic trajectories, a significant computational effort. We found that all three driving protocols, which share the same $H(t),T(t)$, indeed have the same mutual information. As expected, the mutual information lags in response to the changing conditions (Fig.~\ref{fig:KZ_twocell}). As with the other results in this manuscript, the degeneracy between the three protocols is true for both the Sch\"ogl dynamics and the equilibrium  simulations of the companion Landau model (using Metropolis Monte-Carlo).

\emph{Discussion}---In this manuscript we considered a simple theoretical question with a minimal model of cell-to-cell communication: two cells that exchange a molecule and thereby share information, while each cell's parameters change with time. We sought to discern whether cell-to-cell information sharing could be robust to varying individual cell parameters. We showed that indeed there are different single-cell driving protocols that lead to the same system response as long as the collective coordinates have the same dynamics. This suggests a robustness in cell-to-cell communication and a flexibility in the routes cells can take to achieve the same information-sharing outcomes. Despite the minimal nature of the model here studied, our results generalize to all models with either demographic or thermal noise that belong to the Ising mean-field universality class. 

We compared the cell-to-cell stochastic dynamics of the Sch\"ogl model with the companion equilibrium system---an extension of Landau theory, that had been used in the past to describe phase transitions in minerals \cite{salje1992}; surface and hydration forces \cite{kornyshev1992}; and even more generally \cite{kornyshev1995}. To the best of our knowledge, this is the first example in the literature of KZ scaling of 4-parameter nonequilibrium dynamics, here applied in the context of cell-to-cell information. Inspired by the analogy between the systems, one may consider reversing the sign of $g$, making the exchange molecule inhibitory, rather than excitatory, to the other cell. In the Landau formulation, this corresponds to anti-ferromagnetic interactions between the two subsystems. However, in the Schl\"ogl case, one must be careful since a negative $g$ would imply that when there are zero copies of molecule $X$, having nonzero $Y$ could lead to a negative number of $X$ molecules---a contradiction. Therefore, mechanistically, one would need to model an `interaction' where $X$ encounters $Y$.

The dependence of the two-cell system on its low-dimensional representation could potentially allow cells to optimize other aspects of their function without compromising information exchange, a possible mechanism for evolutionary adaptation. Indeed, such degeneracy in single-cell configurations obeying collective coordinates could extend beyond two-cell systems to entire tissues or even larger biological systems.

\emph{Data availability}---All code and data used for this manuscript are freely available to download in \href{https://github.com/AmirErez/TwocellInformationPy}{https://github.com/AmirErez/TwocellInformationPy}.

\emph{Acknowledgments}---Numerical simulations were paid for by AE's startup funds. MV was supported by the Dutch Research Council, NWO Groot grant, number 2019.085.


\bibliographystyle{unsrt}

\clearpage
\appendix
\part*{SI Appendix}

\pagenumbering{arabic}
\renewcommand{\thefigure}{S\arabic{figure}}
\setcounter{figure}{0} 
\renewcommand{\thetable}{S\arabic{table}}
\renewcommand{\thesubsection}{S\arabic{subsection}}
\renewcommand{\theequation}{S\arabic{equation}}
\setcounter{equation}{0}

\section{A: Visualizing the bifurcation transition}
\label{appsec:Pxy}
\begin{figure}[ht]
    \centering
    \includegraphics[width=0.99\columnwidth]{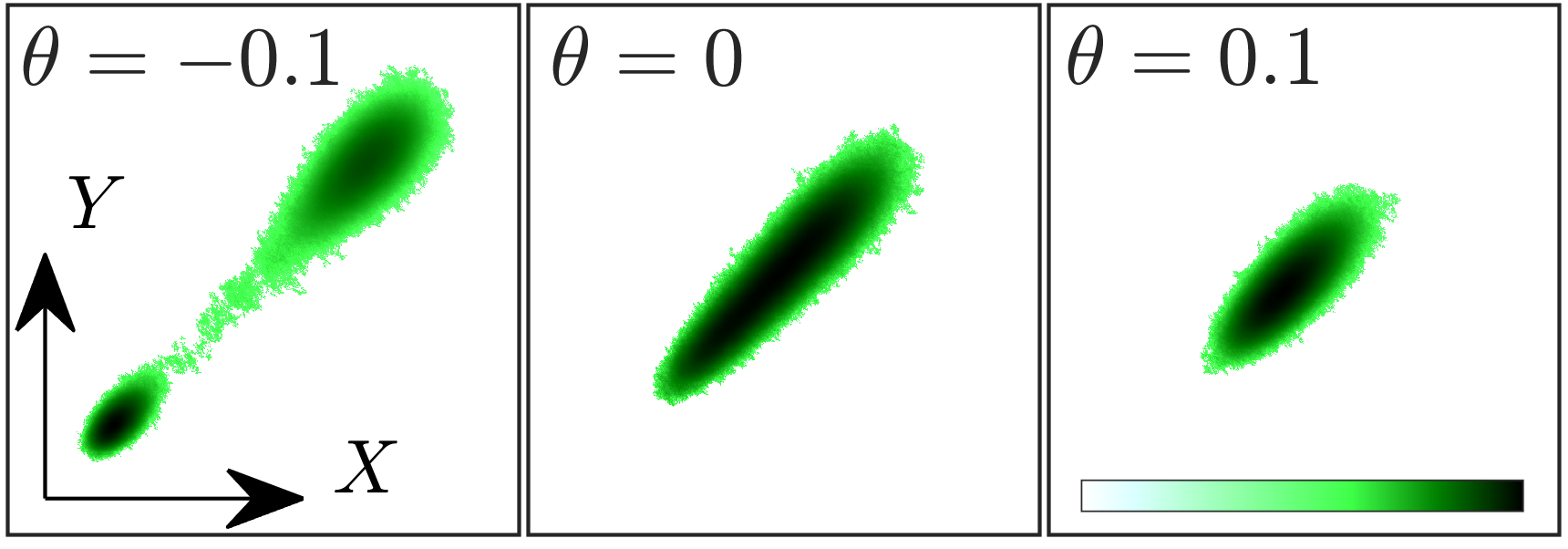}
    \caption{Examples of the joint distribution $P(X,Y)$ shown as a heatmap, calculated from Gillespie simulations of the Sch\"ogl model \cite{erez2020}. The bifurcation point at $\theta=0$ separates a bimodal from a unimodal steady state. Color map corresponds to $\log P$. In all simulations, $h_X=h_Y=0$ and $n_c=1000$.}
    \label{fig:pdfs}
\end{figure}

To make explicit the shape of the probability distributions for the molecule counts of the two cells ($X$,$Y$), we show in Fig.~\ref{fig:pdfs} the effect of the bifurcation parameter, $\theta$, on the two-cell collective state, when $H=0$. Note the flat-topped shape at the critical point \cite{erez2020}.

\section{B: Gaussian mutual information}\label{appsec:GaussianMI}

Linearizing a stochastic differential equation around a deterministic fixed point $\vec{\mu}$ yields an Ornstein-Uhlenbeck process \cite{klebaner2012}:
\begin{equation}
    d\vec{X}_t = \mathcal{J}\left(\vec{X}_t - \vec{\mu} \right)dt + \sigma d\vec{W}_t,
\end{equation}
where the matrix $\mathcal{J}$ is negative definite and $\vec{X}$ and $\vec{W}$ are column vectors. Pathwise solutions may be computed as
\begin{equation}\label{appeq:OUSolution}
    \vec{X}_t = \vec{\mu} + e^{\mathcal{J}t}\left(\vec{X}_0 -\vec{\mu}\right) + \int_0^t e^{\mathcal{J}(t-s)}\sigma d\vec{W}_s.
\end{equation}
Note that, for normally distributed initial conditions, this can be interpreted as a sum of normally distributed variables, so solutions are normal at all times.

The mutual information \cite{shannon1948} between two variables $V_1$ and $V_2$ is
\begin{equation}
    I(V_1,V_2) = \mathbb{E}\left[\log\left(\frac{P(V_1,V_2)}{P(V_1)P(V_2)} \right) \right].
\end{equation}
If $V_1$ and $V_2$ have a joint normal distribution, then this becomes
\begin{equation}
    I(V_1,V_2) = -\frac{1}{2} \log\left(1 - \frac{\text{Cov}(V_1,V_2)^2}{\text{Var}(V_1)\text{Var}(V_2)} \right).
\end{equation}
It suffices to compute the covariance matrix of our process $\vec{X}_t$ in steady state. This can be done by subtracting the mean from Eqn. \ref{appeq:OUSolution}, right multiplying by it by its transpose, taking expectations and using the It{\^o} isometry \cite{klebaner2012}, taking $t\rightarrow \infty$, and finally using integration by parts. The result is that the covariance matrix $\mathcal{C}$ obeys the Lyapunov equation:
\begin{equation}\label{appeq:lyapunov}
    \mathcal{J}\mathcal{C} + \mathcal{C}\mathcal{J}^T = -\sigma \sigma^T.
\end{equation}
This is a linear system of equations in the coefficients of $\mathcal{C}$ that may be solved algebraically.

The matrices $\mathcal{J}$ and $\sigma$ can be calculated at the level of the deterministic dynamics (Eq.~\ref{eq:LandauForm}), so it is insensitive to the distinction between the Landau and Schl{\"o}gl models. We have done this in previous work \cite{erez2020}, where, in the limit $0<|h_X|, |h_Y| \ll g = 1$, we found
\begin{equation}
    \overline{m}_X, \overline{m}_Y \sim \left(\frac{3(h_X + h_Y)}{2} \right)^{1/3} = \overline{m}.
\end{equation}

We start with the Landau dynamics. These take the form
\begin{equation}
    \begin{gathered}
        dm_X = -\Gamma \partial_{m_X}L d\tau +\sqrt{2 \Gamma k_B T_{MC}}\,dW^{(1)}_{\tau}, \\
        dm_Y = -\Gamma \partial_{m_Y}L d\tau +\sqrt{2 \Gamma k_B T_{MC}}\,dW^{(2)}_{\tau}, \\
    \end{gathered}
\end{equation}
where the Wiener processes used here are dimensionless (with variance $\tau$). Linearizing the deterministic dynamics around $(m_X, m_Y) = (\overline{m}, \overline{m})$ and evaluating the noise matrix at that point, we find that 
\begin{equation}
    \label{eq:JsigmaLandau}
    \mathcal{J} = -\Gamma \begin{bmatrix}
        g + \overline{m}^2 & -g \\
        -g & g + \overline{m}^2
    \end{bmatrix}, \quad
    \sigma = \sqrt{2 \Gamma k_B T_{MC}}\, \mathbb{I}_2,
\end{equation}
where $\mathbb{I}_2$ is the $2\times 2$ identity matrix. Substituting Eq.~\ref{eq:JsigmaLandau} into the Lyapunov equation (Eq.~\ref{appeq:lyapunov}) and solving gives
\begin{equation}
    \label{eq:varM}
    \begin{gathered}
        \text{Var}(m_X) = \text{Var}(m_Y) = \frac{k_B T_{MC} (g+\overline{m}^2)}{2g\overline{m}^2+\overline{m}^4}, \\
        \text{Cov}(m_X, m_Y) = \frac{g k_B T_{MC}}{2g\overline{m}^2+\overline{m}^4}.
    \end{gathered}
\end{equation}
Therefore, the mutual information gives
\begin{equation}
    I_{\text{Landau}} = -\frac{1}{2} \log\left(1 - \frac{g^2}{(g+\overline{m}^2)^2}\right).
\end{equation}
Because the covariance matrix was proportional to $T_{MC}$ under this approximation, this Monte-Carlo completely cancels out from the mutual information at the Gaussian limit. Additionally, note that this expression is even in $\overline{m}$, and therefore in $H$.

Now we turn to the case of the Schl{\"o}gl model, which has demographic noise. The dynamics under the chemical Langevin approximation \cite{gillespie2000} can be written as
\begin{equation}
    \begin{aligned}
        dx_t = \left([)k_1^+ - k_1^- x_t + k_2^+ x_t^2 - k_2^- x_t^3 + \gamma_{YX}y_t - \gamma_{XY}x_t\right)dt \\
        + \sqrt{k_1^+ + k_1^- x_t + k_2^+ x_t^2 + k_2^- x_t^3}\,dW_t^{(1)} \\
        - \sqrt{\gamma_{XY} x_t}\,dW_t^{(3)} + \sqrt{\gamma_{YX} y_t}\,dW_t^{(4)}, \\
        dy_t = \left(q_1^+ - q_1^- y_t + q_2^+ y_t^2 - q_2^- y_t^3 - \gamma_{YX}y_t + \gamma_{XY}x_t\right) dt \\
        + \sqrt{q_1^+ + q_1^- y_t + q_2^+ y_t^2 + q_2^- y_t^3}\,dW_t^{(2)} \\
        + \sqrt{\gamma_{XY} x_t}\,dW_t^{(3)} - \sqrt{\gamma_{YX} y_t}\,dW_t^{(4)}.
    \end{aligned}
\end{equation}
In previous work, we inverted the relations between the Landau parameters and the chemical reaction rates \cite{erez2020}. If we have $n_c \gg 1$, $\theta_X = \theta_Y = 0$, the inverse mapping becomes
\begin{equation}
    \begin{gathered}
        q_1^- = k_1^-, \quad \gamma_{XY} = \gamma_{YX} = gk_1^-, \\
        k_1^+ = n_c k_1^-(h_X + 1/3), \quad q_1^+ = n_c k_1^-(h_Y + 1/3), \\
        k_2^- = q_2^- = k_1^-/(3n_c^2), \quad k_2^+ = q_2^+ = k_1^-/n_c.
    \end{gathered} 
\end{equation}
We define the mean reactive/ non-exchange propensity for $X$:
\begin{equation}
    \begin{aligned}
        \overline{R}_X &= k_1^+ + k_1^- \overline{x} + k_2^+ \overline{x}^2 + k_2^- \overline{x}^3, \\
        &= n_c k_1^- \left[h_X + 1/3 + (\overline{m}+1) \right.\\
        &\qquad \qquad \qquad \qquad \left. + (\overline{m}+1)^2 +\frac{(\overline{m}+1)^3}{3} \right] \\
        &= n_c k_1^- \left[h_X + 8/3 + 4 \overline{m} + 2\overline{m}^2 + \frac{\overline{m}^3}{3}\right].
    \end{aligned}
\end{equation}
Analogously for $Y$, we have
\begin{equation}
   \overline{R}_Y = n_c k_1^- \left[h_Y + 8/3 + 4 \overline{m} + 2\overline{m}^2 + \frac{\overline{m}^3}{3}\right]. 
\end{equation}
Finally, we define the mean diffusive propensity:
\begin{equation}
    \overline{D} = \gamma_{XY}\overline{x} = gk_1^-n_c(\overline{m}+1).
\end{equation}
The noise matrix in the linearization is the matrix evaluated at $(x,y) = (n_c(\overline{m}+1), n_c(\overline{m}+1))$. We find that
\begin{equation}
    \sigma = \begin{bmatrix}
        \sqrt{\overline{R}_X} & 0 & -\sqrt{\overline{D}} & \sqrt{\overline{D}} \\
        0 & \sqrt{\overline{R}_Y} & \sqrt{\overline{D}} & -\sqrt{\overline{D}}
    \end{bmatrix}.
\end{equation}
Evaluating the derivative of the deterministic part at the fixed point gives
\begin{equation}
    \mathcal{J} = -k_1^- \begin{bmatrix}
        g + \overline{m}^2 & -g \\
        -g & g + \overline{m}^2
    \end{bmatrix},
\end{equation}
which is proportional to the result from the Landau case. It is readily apparent that $\sigma$ is not even in $\overline{m}$, so we expect asymmetry in $H$ with demographic noise. The case with $h_X = h_Y$ allows some simplification:
\begin{eqnarray}
    I_{\text{Schl\"ogl}} &=& -\frac{1}{2} \log\left(1-\rho^2 \right) \\
    \rho^2 &=&  \frac{4 g^2 \left(-\overline{m}^2+\overline{m}+2\right)^2}{\left(2 g \left(2   \overline{m}^2+\overline{m}+2\right)+\left(\overline{m} \left(\overline{m}+2\right)+4\right)\overline{m}^2\right)^2} \nonumber
\end{eqnarray}

\bigskip
In the general case where $h_X\neq h_Y$, the expression is messy and is studied numerically.

\end{document}